\documentclass[aps,pra,twocolumn,english,showpacs,10pt]{revtex4-1}
\usepackage{amsmath}
\usepackage{amssymb}
\usepackage{graphicx}
\usepackage{babel}
\usepackage{cases}
\usepackage{floatrow}
\usepackage{natbib}

\makeatletter

\begin{document}

\title{Dicke Phase Transition in a Disordered Emitter-Graphene Plasmon System}
\author{Yu-Xiang Zhang}
\email{iyxz@phys.au.dk}
\author{Yuan Zhang}
\email{yzhang@phys.au.dk}
\author{Klaus M{\o}lmer}
\email{moelmer@phys.au.dk}
\affiliation{Department of Physics and Astronomy, Aarhus University, DK-8000 Aarhus C, Denmark}

\begin{abstract}
We study the Dicke phase transition in a disordered system of emitters coupled to the plasmonic modes of
a graphene monolayer. This system has unique properties associated with the tunable, dissipative and
broadband characters of the graphene surface plasmons, as well as the disorder due to the random spatial distribution and the
inhomogeneous line-width broadening of the emitters. We apply the Keldysh functional-integral approach,
and identify a normal phase, a superradiant phase and a spin-glass phase of the system.
The conditions for these phases and their experimental signatures are discussed.
\end{abstract}
\maketitle

The Dicke model \cite{Dicke1954}, which describes the collective
coupling between an ensemble of emitters and a radiation field,
implies a superradiant (SR) phase \cite{Hepp1973,Wang1973} characterized
by a non-zero electromagnetic field excitation and
a collective atomic polarization \cite{Emeljanov1976}.
The validity of the theory predicting the SR phase, especially the proper treatment of $A^2$ \cite{nogo1981}
and $P^2$ terms \cite{Bamba2014}, has been questioned, but has been recently clarified
 \cite{Keeling2007,Vukics2012,Vukics2014,Vukics2015,Grieser2016},
and the SR phase has now been observed experimentally in cold atom systems \cite{Baumann2010,Baumann2011,Brennecke2013,Baden2014,Klinder2015,Zhiqiang2017}
where an effective Dicke model is constructed via cavity-assisted Raman transitions \cite{Dimer2007}.
The Dicke model and its phase transitions have also been extended to scenarios with
multi-mode cavities \cite{Tolkunov2007,Gopalakrishnan2011,Strack2011,Sieberer2016,Buchhold2013},
cavity losses \cite{Buchhold2013,Torre2013,Kirton2017} and
time-dependent couplings \cite{Bastidas2012} as well as other systems
like superconducting circuits
\cite{Viehmann2011,Bamba2016}, Dicke lattice models \cite{Zou2014}, etc.
These proposals display the richness of phenomena associated with the collective and superradiant
light-matter interaction and stimulate studies of the relation between critical behavior and quantum
entanglement \cite{Lambert2004},
quantum chaos \cite{Emary2003} and non-equilibrium dynamics \cite{Sieberer2016}
in a variety of different physical systems.

\begin{figure}[b]
    \includegraphics[width=\textwidth]{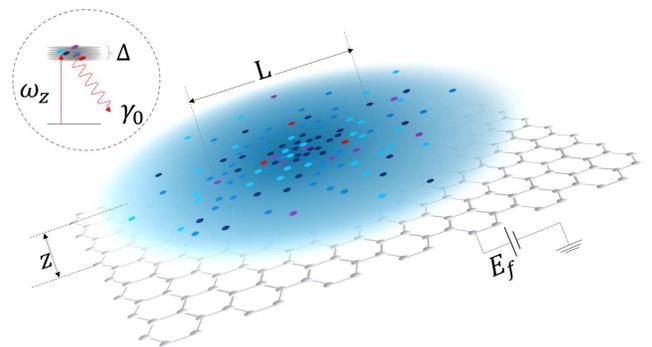}
    \caption{Emitter-graphene system. An ensemble of $N$ emitters with
    spontaneous emission rate $\gamma_0$ and transition frequency inhomogeneously broadened
    by $\Delta$ around a central transition frequency $\omega_z$, is distributed in a layer
    with horizontal dimension $L$ at height $z$.
    The Fermi energy $E_f$ of the graphene electrons can be tuned by gate doping.
    }
    \label{fig-system}
\end{figure}

In this Letter, we investigate the possibility of observing the Dicke SR phase transition
within a system of emitters coupled to surface plasmons (SP).
The SP are evanescent electromagnetic modes
confined near conductor-dielectric interfaces \cite{Pustovit2009}.
Their compressed mode volumes enable strong near-field light-emitter couplings \cite{Tame2013,Toermae2015},
which make quantum plasmonics a promising platform for quantum optical effects \cite{Chang2006,Zhang2006}. Recent developments of two-dimensional plasmonic materials \cite{Basov2016} and, particularly, graphene \cite{Grigorenko2012}, which can be tuned by means of a gate potential
\cite{Grigorenko2012,Koppens2011,Tielrooij2015}, motivate us to study the Dicke phase transition in
systems with graphene SP, cf. Fig. \ref{fig-system}.

The extension of the Dicke model to quantum plasmonics must take into account the broadband SP spectral density \cite{Leggett1987,Orth2010,Winter2014} and the intrinsic Ohmic losses in the graphene. Thus, the quantization of SP is more technical than that of optical cavity modes \cite{Huttner1992,Drezet2017,Philbin2010,Gruner1996,Dung1998}.
Moreover, the fact that the graphene SP wavelengths are shorter
than those of free photons by two orders of magnitudes \cite{Koppens2011},
and could be much shorter than the spatial extent of the emitter ensemble,
makes it impossible to associate a
uniform emitter-field coupling strength as commonly used in the Dicke model.
Finally, emitters such as the rare-earth
ions doped in crystal, have randomly distributed positions and
inhomogeneously broadened transition frequencies. The intrinsic dissipation and disorder
will seriously affect the collective coupling to the SP modes
and hence the conditions for the SR phase transition,
and allow the presence of a quantum spin-glass phase \cite{Edwards1975}.

\emph{Theory-} To describe the disordered emitter-graphene system illustrated in
Fig. \ref{fig-system}, we shall establish a
Keldysh functional-integral approach, which
takes the field losses due to the medium into account \cite{Huttner1992}.

A bosonic field $\mathbf{f}(\mathbf{r},\tilde{\omega})$, with three
Cartesian components ($f_a$), position $\mathbf{r}$,
and frequency $\tilde{\omega}$, can be defined with the
commutators $[f_a(\mathbf{r}_1,\tilde{\omega}_1),f^\dagger_b(\mathbf{r}_2,\tilde{\omega}_2)]
=\delta_{ab}\delta(\mathbf{r}_1-\mathbf{r}_2)\delta(\tilde{\omega}_1-\tilde{\omega}_2)$,
$[f_a,f_b]=0$ and $[f_a^\dagger, f_b^\dagger]=0$, such that the quantized electric
field can be written as \cite{Dung1998,Philbin2010,Gruner1996}:
\begin{equation}\label{e-field}
\begin{aligned}
  \mathbf{E}(\mathbf{r})= i\mu_0
  & \sqrt{\frac{\hbar\epsilon_0}{\pi}}
    \int_{0}^{\infty}d\tilde{\omega}\int d^3\mathbf{r'}\, \\
  \times & \tilde{\omega}^2\sqrt{\Im \epsilon(\mathbf{r'},\tilde{\omega})}
    \mathbf{G}(\mathbf{r},\mathbf{r'},\tilde{\omega})
    \cdot\mathbf{f}(\mathbf{r'},\tilde{\omega})+h.c.,
\end{aligned}
\end{equation}
where $\mathbf{G}(\mathbf{r},\mathbf{r'},\tilde{\omega})$ is
the dyadic Green's tensor,
$\mu_0$ and $\epsilon_0$ are the vacuum susceptibility and permittivity,
$\Im \epsilon$ stands for the imaginary part of the relative permittivity, and $h.c.$ is short for `Hermitian conjugate'.
Equation (\ref{e-field}) resembles the particular solution to
Maxwell's equations associated with a quantized current source $\tilde{\omega}\sqrt{\hbar\epsilon_0\Im\epsilon
(\mathbf{r'},\tilde{\omega})/\pi}\mathbf{f}(\mathbf{r'},\tilde{\omega})$.

The Hamiltonian of the system studied by us can be written as
\begin{equation}\label{hamiltonian}
  H=H_0  +\sum_{i=1}^{N}\bigg[ \frac{1}{2}\hbar\omega_{i,z}\sigma_i^z-
  \sigma_i^x\mathbf{d}_i\cdot\mathbf{E}(\mathbf{r}_i) \bigg],
\end{equation}
where $H_0=\int d^3\mathbf{r'}\int_{0}^{\infty} d\tilde{\omega}\,\hbar\tilde{\omega}
\mathbf{f}^\dagger(\mathbf{r'},\tilde{\omega})\mathbf{f}(\mathbf{r'},\tilde{\omega})$
is the free field Hamiltonian, $\omega_{i,z}$, $\mathbf{d}_i$ and $\mathbf{r}_i$ are the
transition frequency, dipole and
position of the $i^{th}$ emitter. We model the emitters as two-level systems
with Pauli operators $\sigma_x^i$ and $\sigma_z^i$. Notice that here the
rotating-wave approximation is not used.

The Hamiltonian in the form of Eq. (\ref{hamiltonian}) has been widely used
in the literature, and should be interpreted within
the multipolar gauge and the term $\mathbf{E}(\mathbf{r}_i)$ of Eq. (\ref{hamiltonian})
should be understood as $\frac{1}{\epsilon(\mathbf{r}_i)\epsilon_0}\mathbf{D}(\mathbf{r}_i)$,
where $\mathbf{D}(\mathbf{r}_i)$ is the displacement field \cite{Vukics2012,Vukics2014,Vukics2015,Grieser2016}.
Equation (\ref{hamiltonian}) further assumes that the distance between any two emitters
is larger than the size of the atoms, since,
otherwise, a residual instantaneous interatomic potential
must be included in the treatment \cite{Vukics2015,Grieser2016}.
Notice that the experimental observations of the SR phase
transitions are based on
effective Dicke models employing Raman processes
\cite{Dimer2007,Baumann2010,Baumann2011,Brennecke2013,Baden2014,Klinder2015,Zhiqiang2017}.
Our theory can be generalized straightforwardly to the quantum plasmonic version of
these models \cite{Dzsotjan2010}.

The Keldysh functional-integral approach is convenient for the analysis of
open system non-equilibrium dynamics in disordered systems \cite{Sieberer2016}.
To apply it, the Pauli operators representing the two-level emitters are replaced by a
real bosonic variable $\phi_i(t)$ restricted to have
unit length, i.e., $\phi_i^2(t)=1$ \cite{Sieberer2016}:
\begin{equation}\label{spin-field}
  \sigma^x_i(t)\rightarrow \phi_i(t),\quad
  \sigma^z_i(t)\rightarrow \frac{2}{\omega_{i,z}^2}(\partial_t \phi_i)^2-1.
\end{equation}
This mapping originates from the correspondence between the
energy gap of quantum models and the correlation length along the `time'
direction of their classical counterparts, and works well for phase transitions
\cite{Strack2011,Torre2013,Sieberer2016,Buchhold2013},
see Refs. \cite{Kogut1979,Ye1993,Kennett2001,Sachdev2011} for further details.
The Keldysh action of the free emitters derived from Eq. (\ref{hamiltonian}) is then expressed as
\begin{equation}\label{action-e}
  S_e=-\sum_{\substack{i=1 \\ a=\pm}}^{N}\int_{C_a}dt
  \bigg[\frac{1}{\omega_{i,z}}(\partial_t \phi_{i,a})^2+
  \lambda_{i,a}(t)(\phi_{i,a}^2-1)\bigg],
\end{equation}
where $\lambda_{i,a}$ is the Lagrange multiplier
introduced for the restriction $\phi_{i,a}^2=1$, and the variables labelled by $a=\pm$ are defined along the
time-integral contours $C_\pm= \mp\infty \rightarrow \pm\infty$
(for steady states, we do not need to specify initial states \cite{Aoki2014}).

In the Keldysh functional integral approach, we can formally integrate out the
degrees of freedom of $\mathbf{f}(\mathbf{r},\tilde{\omega})$
and get the Keldysh action for the emitter-emitter coupling mediated by them \cite{sp}:
\begin{equation}
\begin{aligned}\label{actionp}
 & S_{ee}^{(p)}=\sum_{i,j=1}^{N}
  \int_{-\infty}^{\infty} \frac{d\omega}{2\pi}
    \begin{pmatrix}
      \phi_{i,c} & \phi_{i,q}
    \end{pmatrix}_{-\omega}\\
    &\qquad\qquad\quad \times \begin{pmatrix}
                             0 & h_{ij}^*(\omega) \\
                             h_{ij}(\omega) & 2i\Im h_{ij}(|\omega|)
                           \end{pmatrix}
                           \begin{pmatrix}
                             \phi_{j,c} \\
                             \phi_{j,q}
                           \end{pmatrix}_\omega,
\end{aligned}
\end{equation}
where the $\omega$-dependent coupling strength is
\begin{equation}\label{hij}
 h_{ij}(\omega)= \frac{\omega^2}{2\hbar\epsilon_0c^2}
  \mathbf{d}_i\cdot\mathbf{G}(\mathbf{r}_i,\mathbf{r}_j,\omega)\cdot\mathbf{d}_j.
\end{equation}
Note that we have passed to the Fourier domain with frequency variable $\omega$,
and have transformed to the so-called `classical' (`quantum') fields $\phi_{i,c(q)}$
by the Keldysh rotation $\phi_{i,c(q)}=[\phi_{i,+}+(-)\phi_{i,-}]/\sqrt{2}$ \cite{Sieberer2016}.
The corresponding transformation of the Lagrange multipliers $\lambda_{i,c(q)}$ is
$\lambda_{i,c(q)}=\lambda_{i,+}+(-) \lambda_{i,-}$.

\emph{Spatial Disorder-}To treat the disorder in the emitter system, we follow the strategy of random-bond models
widely used in the studies of spin-glasses \cite{Fischer1993}.
That is, the real and imaginary parts of the coupling strength,
$\{\Re h_{ij}(\omega), \Im h_{ij}(\omega)\}_{i\neq j}$, which are functionals of
the emitter positions and dipoles, are viewed as random variables following a
multi-component Gaussian distribution (neglecting higher order moments)
with the mean and the covariance given by
\begin{subequations}
\begin{align}
\begin{split}\label{meanm}
&\overline{h}_{(2)}(\omega)=\int d^3\mathbf{r}_a d^3\mathbf{r}_b
p(\mathbf{r}_a,\mathbf{r}_b)h_{ab}(\omega),
\end{split}\\
\begin{split}\label{covm}
  & M(\omega, \omega')=\int d^3\mathbf{r}_a d^3\mathbf{r}_b p(\mathbf{r}_a,\mathbf{r}_b)\\
& \quad\times\begin{pmatrix}
    \delta\Re h_{ab}(\omega)\delta \Re h_{ab}(\omega') &
    \delta\Re h_{ab}(\omega)\delta \Im h_{ab}(\omega') \\
    \delta\Im h_{ab}(\omega)\delta \Re h_{ab}(\omega')
    &
    \delta\Im h_{ab}(\omega)\delta \Im h_{ab}(\omega')
  \end{pmatrix},
\end{split}
\end{align}
\end{subequations}
where $p(\mathbf{r}_a,\mathbf{r}_b)$ denotes the probability distribution of the positions of two emitters
(the average over $\{\mathbf{d}_i\}$
is implicitly assumed), and `$\delta$' denotes the difference with respect to the mean
value of the real and imaginary parts of $\overline{h}_{(2)}(\omega)$.
For the emitter-graphene system to be investigated later,
the individual terms $h_{ii}(\omega)$ are identical for all $i$, since they are determined only by
the height $z$ of the emitter layer over the graphene.
We shall denote their values as $\overline{h}_{(1)}(\omega)$.

\emph{Inhomogeneous Broadening-}Emitters such as rare-earth ions doped in crystals
experience inhomogeneous broadening of their transition spectrum,
cf. Fig. \ref{fig-system}. To take this into account,
the conventional method is to divide the ensemble into groups of emitters with
same transition frequency \cite{Goto2008,Zou2014}. Here, we do not follow
this method but rather assume
the transition frequency $\omega_{i,z}$ follows a
Gaussian distribution centered at $\omega_z$ with standard deviation $\Delta$.
Thus, the broadening can be treated statistically and contributes a new term to the
Keldysh action of the system
\begin{equation}\label{sb}
  S^{(b)}= i\frac{\Delta^2}{2\omega_z^4}\sum_{i=1}^{N}\bigg(\int d\omega
     \omega^2\phi_{i,c}(-\omega)\phi_{i,q}(\omega)\bigg)^2.
\end{equation}
In Ref. \cite{sp} we show that the main effect of
$S^{(b)}$ is to shift
the covariance $M(\omega,\omega')$ defined in Eq. (\ref{covm})
by terms that scale as $(\Delta \frac{\omega\omega'}{\omega_z^2})^2$
and are negligible for a large $N$.

\emph{Order Parameters-}To distinguish the different phases of the system, we introduce the following order parameters
\cite{Buchhold2013,Kennett2001,Sachdev2011,Torre2013,Sieberer2016,Fischer1993}:
\begin{equation}\label{Q}
\begin{aligned}
  Q_{\alpha\beta}(\omega,\omega')& =-i\frac{1}{N}\sum_{i=1}^{N}\langle
        \phi_{i,\alpha}(\omega)\phi_{i,\beta}(\omega')\rangle, \\
  \psi_\alpha(\omega) & = -\frac{1}{N}\sum_{i=1}^{N}
    \langle \phi_{i,\alpha}(\omega)\rangle,
  \end{aligned}
\end{equation}
where $\alpha,\beta\in \{c, q\}$.
$Q_{cq},\, Q_{qc}$ and $Q_{cc}$ are
the retarded, advanced and Keldysh Green's functions of the
emitters \cite{Sieberer2016}, respectively.
$\psi_{c}$ is the average polarization
of the emitters.
For the steady state, we substitute the ansatz that
$\psi_\alpha(\omega)=2\pi\delta(\omega)\psi_\alpha$,
$\lambda_{i,\alpha}(\omega)=2\pi\delta(\omega)\lambda_{i,\alpha}$,
$Q_{\alpha\beta}(\omega,\omega')=2\pi\delta(\omega+\omega')Q_{\alpha\beta}(\omega)$
and introduce the Edward-Anderson order parameter $q_{EA}$
\cite{Edwards1975,Fischer1993,Sieberer2016,Ye1993,Buchhold2013,Strack2011}
to pin down the spin-glass phase:

\begin{equation}\label{qea}
  Q_{cc}(\omega)=Q_{cc}^{reg}(\omega)-i2\pi q_{EA}\delta(\omega),
\end{equation}
where `reg' labels the regular part. In the time domain, we have
$q_{EA}\propto\lim_{t\rightarrow\infty}\frac{1}{N}\sum_i\langle\sigma^x_i(t)\sigma^x_i(0)\rangle$.
Thus a finite $q_{EA}$ implies an infinite correlation time of the
individual emitter dipoles.

The steady state of the system and the values of the order parameters
are determined by the saddle-point equations of the Keldysh action \cite{sp}.
This leads to the identification of three different
phases: the superradiant (SR) phase with
($q_{EA}\neq 0, \psi_c\neq 0$),
the spin-glass (SG) phase with ($q_{EA}\neq 0, \psi_c=0$), and the normal phase with
($q_{EA}=0, \psi_c=0)$.

\emph{Results-}
We model the system depicted in Fig. \ref{fig-system}, as
a layer of $N$ emitters positioned at a distance $z$ over the graphene monolayer.
The emitter dipoles $\{\mathbf{d}_i\}_i$ are aligned to be
perpendicular to the graphene layer and their magnitudes
are quantified by the spontaneous emission rate $\gamma_0$.
The graphene is modeled as a two dimensional surface with
conductivity $\sigma(E_f,\tau;\omega)$ \cite{Hanson2008} given
in the local random phase approximation \cite{Koppens2011},
where $E_f$ is the Fermi energy tunable by gate doping and
$\tau$ is the relaxation time accounting for the electron-phonon scattering (we use $\tau=10^{-13}s$ \cite{Koppens2011}).
The in-plane positions of the emitters are assumed to
follow independent Gaussian distributions
with width $L$. Our results thus depend on the set of parameters
$N, L, z, E_f, \omega_z, \gamma_0, \Delta$.
To focus on the phase transitions associated with the
plasmonic evanescent modes, we shall
omit the weak coupling to the propagating modes \cite{Keeling2007,Vukics2015} in the following.
This is done by replacing the total dyadic Green's function by its `scattering'
part which contains the information of the graphene SP \cite{sp}.

\begin{figure}[b]
\centering
\includegraphics[width=\textwidth]{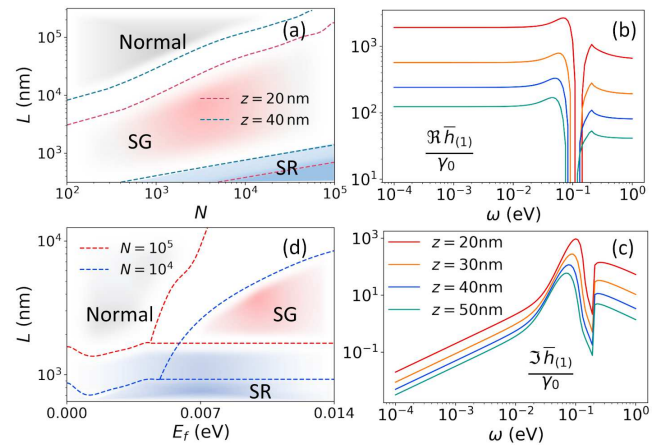}
\caption{Results for systems with $\omega_z=0.5\,\mathrm{eV}$ and $E_f=0.1\,\mathrm{eV}$: Panel (a) shows the $N-L$
phase diagram for $\gamma_0=10^{-5}\,\mathrm{eV}$, with different Normal-SG and SG-SR boundaries for
$z= 20 \mathrm{nm}$(red, lower dashed curves) and $40 \mathrm{nm}$(green, upper dashed curves); Panel
(b) shows the value of $\Re\overline{h}_{(1)}(\omega)$, the energy shift induced by graphene, and panel
(c) shows the value of $\Im\overline{h}_{(1)}(\omega)$, the graphene-induced emitter damping, as functions of frequency
and different heights, $z=$20 nm(red, top), 30 nm(orange), 40 nm(blue), 50 nm(green, bottom);
Panel (d) shows the $E_f-L$ phase diagram for $z=50 \mathrm{nm}, \gamma_0=10^{-8} \mathrm{eV}$, with different
Normal-SG and SG-SR boundaries for $N=10^4$(blue, lower dashed curves) and $10^5$(red, upper dashed curves).}
\label{fig2}
\end{figure}

Fig. \ref{fig2}(a)  shows the location of the phase transitions
as a function of the ensemble size and number of emitters. It demonstrates that
the SR phase favors higher emitter densities.
We also find that the phase diagram changes only little due to
inhomogeneous broadening: For $z=20\,\mathrm{nm}$ and N=100 the
Normal-SG phase boundary shifts $L$ downward by only about 60 nm for a
broadening as large as $\Delta=0.1\,\mathrm{eV}$ (here and throughout, $\hbar=1$).

Although smaller $z$ implies stronger emitter-graphene SP couplings,
Fig. \ref{fig2}(a) shows that when the emitters are moved from
the $z=40\,\mathrm{nm}$
to $z=20\,\mathrm{nm}$ distance to the graphene, the
Normal-SG phases and SG-SR phase boundaries shift downward, i.e., they occur
for higher emitter densities.
When $z$ is decreased, there is a complicated interplay between the enhanced SP-induced energy shift,
see Fig. \ref{fig2}(b), leading to the Dicke SR phase, and the increased damping
of the emitters, due to the same coupling, see Fig. \ref{fig2}(c).
The competition between these effects is the main cause for the shift in
the phase-transition boundaries.
We note, however, for extremely small $z$, emitter-graphene
bound states may form
\cite{Tong2010,Yang2017,Thanopulos2017,Gonzalez-Tudela2014,Gonzalez-Tudela2010}
so that different behavior,
including polarization of the emitters, should be expected.

One may try to understand the SR phase of our system by comparing it with
the Dicke model of a single cavity mode, where the effective emitter-emitter
coupling Hamiltonian is given by
$H_{eff}=-\sum_{i,j}J\sigma_i^x\cdot\sigma_j^x$,
$J=g^2\omega_c/(\omega_c^2-\omega^2)$ \cite{Strack2011}, and the SR phase
is reached when $g^2N > \omega_z\omega_c/4$. In our model, $\Re h_{ij}(\omega)$
plays the role of $J$ and the mean $\Re\overline{h}_{(2)}(\omega)$
does not meet the equivalent SR criterion.
However, smaller size sub-ensembles of emitters might
experience strong enough mutual coupling. This fact
is indicated by the large fluctuations of $\Re h_{ij}(\omega)$ resulting from the disorders,
which are shown in Fig. \ref{sp-figure}(e-f) of Ref. \cite{sp}.
Such sub-ensembles would contribute significantly to the
averaged polarization $\psi_c$ of the system of emitters and lead
to the SR phase. To properly account for the role of such sub-ensembles, a
more refined description than the current mean-field approach will be required.
A similar, so far un-noticed, relaxation of the SR criterion on the average coupling strength occurs for the Dicke model with a multi-mode cavity \cite{Strack2011}.

\begin{figure}[b]
\centering
\includegraphics[width=\textwidth]{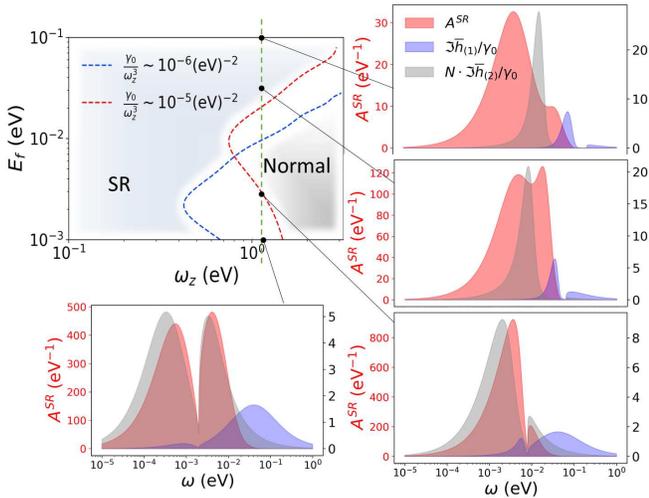}
\caption{$E_f-\omega_z$ phase
diagram for a system with $z=50\,\mathrm{nm}$, $L=10^3\,\mathrm{nm}$, $N=2\times 10^4$, and two different values of
$\gamma_0/\omega_z^3$, so that when
$\omega_z=0.5\,\mathrm{eV}$, $\gamma_0=10^{-7}$ (blue dashed)
or $10^{-6}\,\mathrm{eV}$ (red dashed). For the case of $\omega_z=1\,\mathrm{eV}$, the spectral densities
$A^{SR}=-2\Im Q_{cq}(\omega)$ (red shaded), $\Im\overline{h}_{(1)}$ (grey shaded)
and $\Im\overline{h}_{(2)}$ (blue shaded) are shown
for different Fermi-energies, $E_f=0.1,\,0.032,\, 0.004,\,0.001\,\mathrm{eV}$. The values of $\Im\overline{h}_{(1,2)}$
are shown on the right hand vertical axes.}
\label{fig3}
\end{figure}

In the following we discuss the effect of tuning the Fermi energy $E_f$, a possibility unique to graphene.
The SG-SR phase boundary is insensitive to $E_f$ \cite{sp}.
A higher $E_f$, however, leads to stronger graphene SP-induced
emitter-emitter coupling \cite{Tielrooij2015,Koppens2011} and
facilitates the Normal-SG phase transition
as shown in the phase diagram of Fig. \ref{fig2}(d).
It also shows a triple point and the Normal-SR phase boundary
which are absent in Fig. \ref{fig2}(a). However, there is
also a subtle SR$\rightarrow$Normal$\rightarrow$SR transition
with an increasing $E_f$.

To understand it, we borrow ideas from the studies of spin-boson models
\cite{Leggett1987,Orth2010,Winter2014}, which suggest that the following three quantities might be pertinent:
the emitter spectral
response yield from the emitter linear susceptibility, $A^{SR}(\omega)=-2\Im Q_{cq}$;
the spectral density $\Im \overline{h}_{(1)}(\omega)$
and the `many spin' extension of the spectral density, $\Im\overline{h}_{(2)}(\omega)$.
The spectral density is the central concept of models where a single spin couples to
a continuum of bosons \cite{Leggett1987}.
We note that only $A^{SR}(\omega)$ depends on $\omega_z$ \cite{Strack2011} while
$\Im \overline{h}_{(1)}(\omega)$ and $\Im \overline{h}_{(2)}(\omega)$ depend
on the magnitude of the emitter dipoles quantified by $\gamma_0/\omega_z^3$.

To look closer at the Normal-SR transition,
we depict an $E_f-\omega_z$ phase diagram in Fig. \ref{fig3},
for different values of  $\gamma_0/\omega_z^3$. The frequency dependence of $A^{SR}$,
$\Im \overline{h}_{(1)}$ and $\Im \overline{h}_{(2)}$ are shown in Fig. \ref{fig3}
for the four different Fermi energies $E_f=0.1,\,0.032,\, 0.004,\,0.001\,\mathrm{eV}$.
There are gaps between the positions of the
peaks of $\Im \overline{h}_{(1)}(\omega)$ and those of $\Im \overline{h}_{(2)}(\omega)$,
because the `short-range' modes, important for self-interaction term
$\Im \overline{h}_{(1)}(\omega)$, cannot propagate far enough to affect the
averaged emitter-emitter coupling.
Changing $E_f$ shifts the peaks of $A^{SR}$, $\Im \overline{h}_{(1)}$
and $\Im \overline{h}_{(2)}$, and we observe a closer overlap of $A^{SR}(\omega)$
with $\Im \overline{h}_{(1)}(\omega)$, reflecting the influence of
the SP-induced atomic decay, when the
system is closer to the regime of the Normal phase.
For the number of emitters $N$ applied here, $\Im\overline{h}_{(1)}(\omega)$
and $N\Im \overline{h}_{(2)}(\omega)$ are comparable and suggest that the subtle
$E_f$-dependence of the phase transition observed in Fig. \ref{fig2}(d)
and Fig. \ref{fig3} is a finite-N effect
relevant to the graphene SP-induced emitter decay.

Additionally, the peaks of $A^{SR}$ and $\Im\overline{h}_{(1,2)}$ shown in
Fig. \ref{fig3} generally occur far from the emitter resonance $\omega_z$. This indicates that
the influence of the inhomogeneous broadening, which scales as
$(\Delta \frac{\omega\omega'}{\omega_z^2})^2$, is small. Moreover,
their marked frequency dependence invalidates the Markov approximation, which would
replace $\Im\overline{h}_{1,2}(\omega)$
by a constant taken at the atomic transition energy \cite{Dung2002,Gonzalez-Tudela2014,Vega2017,Gonzalez-Tudela2010}.
Indeed, our formalism considers
the full spectral dependencies and does not apply the Markov approximation.

\emph{Summary and Outlook-}
To summarize, applying the Keldysh functional-integral
approach, we have studied the Dicke phase
transitions between the superradiance phase, spin-glass phase and the normal phase in a disordered emitter-graphene surface plasmon
system. Our formalism is a generalization of the spin-boson
model \cite{Leggett1987} to the many-spin system and is valid for general plasmonic systems.
The variety of nanoscale plasmonic systems, and especially 2D materials like the graphene monolayer,
constitute excellent platforms to test the fundamental collective
phenomena of the Dicke model, and its effects in
quantum optics, non-equilibrium
dynamics of driven dissipative system,
and condensed matter physics.

The superradiant phase is characterized by the emitter polarization.
The spin-glass phase behaves differently from the superradiant phase
at the low frequency regime of the emitter spectral response $-2\Im Q_{cq}(\omega)$
\cite{Buchhold2013,Torre2013,Sieberer2016}. Thus they may be distinguished by
observing their radio-frequency spectral response \cite{Stewart2008,Haussmann2009}.
Here we considered only the plasmonic evanescent modes, and disregarded weakly coupled optical scattering modes from the analysis. By employing an optical cavity, it may be possible to observe a hybrid coupling of the emitters to both surfacce plasmons and a cavity mode, and to use the cavity response and transmission spectrum, as a signature of the surface plasmon Dicke phase transition \cite{Buchhold2013,Dimer2007}.

\emph{Acknowledgement-}
We sincerely thank Frank Koppens, Klass-Jan Tielrooij, Daniel Cano Reol, and Darrick Chang
for useful discussions and suggestions.
This work was supported by
European Union’s Horizon 2020 research and innovation
program (No. 712721, NanoQtech) and the Villum Foundation.

\section{Supplemental Material}
\numberwithin{figure}{section}
\numberwithin{equation}{section}

In this supplemental material, we shall present details of our
derivations of the Dicke phase transitions
based on the Keldysh functional-integral approach.
The Keldysh action of a system with Hamiltonian $H$
and Lindblad dissipation operator $\{L_\alpha\}$
is formulated, by representing the dynamical variables by $\psi$,
according to\cite{Sieberer2016}
\begin{equation}\label{keldysh action}
\begin{aligned}
  S &=\int dt\;\bigg[ \psi_+^*i\partial_t\psi_+-\psi_-i\partial_t\psi_{-}-(H_+-H_-)\\
  &-   i\sum_\alpha \gamma_\alpha( L_{\alpha+}L_{\alpha-}^*-
  \frac{1}{2}L_{\alpha+}^*L_{\alpha+}-\frac{1}{2}L_{\alpha-}^*L_{\alpha-})\bigg].
\end{aligned}
\end{equation}
We employ two sets of bosonic variables,
$\phi_i$ for the emitters and $\mathbf{f}(\mathbf{r}',\tilde{\omega})$
for the electromagnetic environment.

This Supplemental Material includes the following contents:
\begin{description}
   \item[\ref{secA}] The Keldysh action of the free emitters;
  \item[\ref{secB}] The Keldysh action of the SP field, emitter-field coupling, and effective
  emitter-emitter coupling mediated by the graphene SP field;
  \item[\ref{secC}] Averaging over the disorders;
  \item[\ref{secD}] Determining the phases by the saddle point equations;
  \item[\ref{IB}] The treatment of inhomogeneous broadening;
  \item[\ref{secF}] Some notes about the calculation of the emitter-graphene surface plasmons system,
  including curves of the averaged coupling strength and the covariance matrix elements.
\end{description}

\subsection{Action of the Free Emitters}\label{secA}
The Keldysh action for the free emitters is given as Eq.
(\ref{action-e}) in the main text and is derived from Eq. (\ref{keldysh action})
with the mapping
\begin{equation}
  \sigma^x_i(t)\rightarrow \phi_i(t),\quad
  \sigma^z_i(t)\rightarrow \frac{2}{\omega_z^2}(\partial_t \phi_i)^2-1.
\end{equation}
where we have omitted the effect of inhomogeneous broadening.
Discussion on that is deferred to Sec. \ref{IB}.
Then we substitute $\phi_i(t)$ into Eq. (\ref{keldysh action}).
Since $\phi_i(t)$ is a real variable,
the first two terms of Eq. (\ref{keldysh action}) are time-derivative terms,
that is, $\phi_\pm^* i\partial_t\phi_\pm=\phi_\pm i\partial_t\phi_\pm=\frac{1}{2}i\partial_{t}(\phi_\pm^2)$.
These terms are negligible because they have no effects on the action
after the integral over time.

The restriction $\phi^2(t)=1$ is imposed by multiplying the
Keldysh partition function by the delta functions
$\prod_t \delta(\phi^2_\pm(t)-1)$. This process brings
Lagrange multipliers $\lambda_\pm(t)$ to the action
according to the relation that:
\begin{equation}
  \prod_t\delta(\phi^2_\pm(t)-1)=
  \int D\lambda_\pm(t)e^{i\int dt\;\lambda_\pm(t)(\phi_\pm^2(t)-1)}.
\end{equation}

Then, we perform the Keldysh rotation, a unitary transformation
of the contour index:
\begin{equation}
\begin{aligned}
  & \phi_c=\frac{1}{\sqrt{2}}(\phi_+ +\phi_-),\qquad
  \phi_q=\frac{1}{\sqrt{2}}(\phi_+ -\phi_-),\\
  & \lambda_c=\lambda_+ +\lambda_-,\qquad
  \lambda_q=\lambda_+ -\lambda_-.
\end{aligned}
\end{equation}
where the subscripts 'c' and 'q' stand for 'classical' and 'quantum',
respectively \cite{Sieberer2016}.
The constraint equation then amounts to inclusion of the Lagrange multiplier term
\begin{equation}
  2\int_t\lambda_c(t)\phi_c(t)\phi_q(t)+\lambda_q(t)(\phi_c^2(t)+\phi_q^2(t)-2),
\end{equation}
into the action, where $\int_t$ is shorthand for $\int dt$.
Retaining only its static contribution, we use the ansatz that
\begin{equation}
  \lambda_{i,\alpha}(\omega)=2\pi\lambda_{i,\alpha}\delta(\omega)
\end{equation}
in the Fourier domain, where $\alpha\in\{c,q\}$.
Finally, the Keldysh action of the free emitters
is written as
\begin{equation}
\begin{aligned}
  S_e= & \sum_{i=1}^{N}\int_w(\phi_{i,c},\phi_{i,q})_{-\omega}
  \begin{pmatrix}
    \lambda_{i,q} & \lambda_{i,c}-\frac{\omega^2}{\omega_z}\\
    \lambda_{i,c}-\frac{\omega^2}{\omega_z} & \lambda_{i,q}
  \end{pmatrix}
  \begin{pmatrix}
    \phi_{i,c} \\
    \phi_{i,q}
  \end{pmatrix}_\omega\\
  &    -2\sum_{i=1}^{N}\lambda_{i,q}2\pi\delta(0),
\end{aligned}
\end{equation}
where $\int_w$ is shorthand for $\int\frac{d\omega}{2\pi}$.

\subsection{Action for the Plasmonic Environment}\label{secB}
The plasmonic electromagnetic environment is quantized through the
complex field $\mathbf{f}(\mathbf{r}',\tilde{\omega})$. Here we denote
it as $f_{a,\mathbf{r},\tilde{\omega}}$,
where ``$a$'' labels the three Cartesian directions.
The Keldysh action of the free plasmonic environment and
its coupling to the emitters is
\begin{equation}\label{action-f-w}
\begin{aligned}
   S_{f,ef} &= \sum_{a}\int_{\tilde{\omega},\mathbf{r}',\omega} \begin{pmatrix}
                               f^*_{a,\mathbf{r},\tilde{\omega};c} &
                               f^*_{a,\mathbf{r},\tilde{\omega};q}
                             \end{pmatrix}_\omega D_{\tilde{\omega}}(\omega)
                             \begin{pmatrix}
                               f_{a,\mathbf{r},\tilde{\omega};c} \\
                               f_{a,\mathbf{r},\tilde{\omega};q}
                             \end{pmatrix}_\omega \\
                            & -
  \sum_{i=1}^N\sum_{a}\int_{\tilde{\omega},\mathbf{r}',\omega}
              g_{ia}(\mathbf{r}',\tilde{\omega})
              \bigg(\phi_{i,-\omega;c}f_{a,\mathbf{r}',\tilde{\omega};q}(\omega)
              \\ & +\phi_{i,-\omega;q}f_{a,\mathbf{r},\tilde{\omega};c}(\omega)\bigg) +
                            g^*_{ia}(\mathbf{r}',\tilde{\omega})\bigg(
                             \phi_{i,\omega;c}f^*_{a,\mathbf{r}',\tilde{\omega};q}(\omega) \\
                             &
                             +\phi_{i,\omega;q}f^*_{a,\mathbf{r}',\tilde{\omega};c}(\omega)\bigg),
\end{aligned}
\end{equation}
where $\int_{\tilde{\omega}}$ is shorthand for
$\int_{0}^{\infty}\frac{d\tilde{\omega}}{2\pi}$, $\int_{\mathbf{r}'}$
is shorthand for $\int d^3\mathbf{r}'$, and
the matrix $D_{\tilde{\omega}}(\omega)$ is defined as
\begin{equation}
  D_{\tilde{\omega}}(\omega)=\begin{pmatrix}
    0 & \omega-\tilde{\omega}-i\epsilon \\
    \omega-\tilde{\omega}+i\epsilon & 2i\epsilon
  \end{pmatrix},
\end{equation}
and $\epsilon$ stands for an infinitesimal positive constant;
the coupling strength is
\begin{equation}
  g_{ia}(\mathbf{r}',\tilde{\omega})=-i\sqrt{\frac{\epsilon_I(
  \mathbf{r}',\tilde{\omega})}{\hbar\pi\epsilon_0}}\frac{\tilde{\omega}^2}{c^2}
  \sum_{b}{\mathbf{d}_i}_{b} \mathbf{G}_{ba}
  (\mathbf{r}_i, \mathbf{r}',\tilde{\omega}).
\end{equation}
In Eq. (\ref{action-f-w}) all terms with identical
indices of $\tilde{\omega}$ and $\omega$
share the same matrix $D_{\tilde{\omega}}(\omega)$. Therefore,
after integrating out the field of $\mathbf{f}(\mathbf{r},\tilde{\omega})$, $S_{f,ef}$ turns
out to be an effective emitter-emitter coupling action:
\begin{equation}
\begin{aligned}
  S^{(p)}_{ee}=-\sum_{i,j=1}^{N}& \int_{\tilde{\omega},\omega}
     \tilde{g}_{ij}(\tilde{\omega})\begin{pmatrix}
                    \phi_{i,c} & \phi_{i,q}
                  \end{pmatrix}_{-\omega}\\
                  & \times \sigma_x D_{\tilde{\omega}}^{-1}\sigma_x
                  \begin{pmatrix}
                    \phi_{j,c} \\
                    \phi_{j,q}
                  \end{pmatrix}_\omega,
\end{aligned}
\end{equation}
where the coupling strength $\tilde{g}_{ij}(\tilde{\omega})$ is
\begin{equation}
\begin{aligned}
  \tilde{g}_{ij}(\tilde{\omega})= &
  \sum_{a}\int_{\mathbf{r}'}g_{ia}(r',\tilde{\omega})g_{ja}^*(\mathbf{r}',\tilde{\omega})\\
  = &  \frac{1}{\pi\epsilon_0\hbar c^2}\tilde{\omega}^2\mathbf{d}_i\cdot
    \Im\mathbf{G}(\mathbf{r}_i, \mathbf{r}_j, \tilde{\omega})\cdot\mathbf{d}_j,
\end{aligned}
\end{equation}
and $\sigma_x$ is the matrix $\begin{pmatrix}
                                0 & 1 \\
                                1 & 0
                              \end{pmatrix}$.
In the derivation of $\tilde{g}_{ij}$, we have used the relation
\begin{equation}
\begin{aligned}
  & \sum_b\frac{\omega^2}{c^2}\int_{\mathbf{r}'}
  \epsilon_I(\mathbf{r}',\omega)\mathbf{G}_{ab}(\mathbf{r}_i,
  \mathbf{r}',\omega)\mathbf{G}^*_{cb}(\mathbf{r}_j, \mathbf{r}', \omega)\\
  = & \Im\mathbf{G}_{ac}(\mathbf{r}_i,\mathbf{r}_j,\omega).
\end{aligned}
\end{equation}
The inverse of $D_{\tilde{\omega}}(\omega)$  is expressed as
\begin{equation}\label{D-1}
  D_{\tilde{\omega}}^{-1}(\omega)=\begin{pmatrix}
               \frac{-2i\epsilon}{(\omega-\tilde{\omega})^2+\epsilon^2} &
               \frac{1}{\omega-\tilde{\omega}+i\epsilon} \\
               \frac{1}{\omega-\tilde{\omega}-i\epsilon} & 0
             \end{pmatrix}.
\end{equation}
Then, using the relations
\begin{equation}
\begin{aligned}
  & \lim_{\epsilon\rightarrow
  0^+}\frac{\epsilon}{(\omega-\omega_\mu)^2+\epsilon^2}=\pi\delta(\omega-\omega_\mu),\\
  & \lim_{\epsilon\rightarrow 0^+}\frac{1}{\omega-\omega_\mu\pm i\epsilon}=
 \mathcal{P}\frac{1}{\omega-\omega_\mu}\mp i\pi\delta(\omega-\omega_\mu),
\end{aligned}
\end{equation}
we can implement the integral of $\tilde{\omega}$ in $S^{(p)}_{ee}$, i.e.,
\begin{equation}
\Lambda(\omega)=\int_{\tilde{\omega}} \tilde{g}_{ij}(\tilde{\omega})\sigma_x
D^{-1}_{\tilde{\omega}}(\omega)\sigma_x.
\end{equation}
The result is
\begin{equation}
  \Lambda(\omega)=\begin{pmatrix}
  0 & F_{ij}(\omega)+i\pi\Delta_{ij}(\omega) \\
  F_{ij}(\omega)-i\pi\Delta_{ij}(\omega) & -2i\pi\Delta_{ij}(\omega),
  \end{pmatrix}
\end{equation}
where the elements of the matrix are
\begin{subequations}
\begin{align}
  & F_{ij}(\omega)=\int_{\tilde{\omega}} \frac{\tilde{\omega}^2}{\pi\epsilon_0\hbar c^2}
  \mathbf{d}_i\cdot\Im\mathbf{G}(\mathbf{r}_i, \mathbf{r}_j, \tilde{\omega})\cdot\mathbf{d}_j\;\mathcal{P}\frac{1}{\omega-\tilde{\omega}},\\
  & \Delta_{ij}(\omega)=\int_{\tilde{\omega}} \frac{\tilde{\omega}^2}{\pi\epsilon_0\hbar c^2}
  \mathbf{d}_i\cdot\Im\mathbf{G}(\mathbf{r}_i, \mathbf{r}_j,
  \tilde{\omega})\cdot\mathbf{d}_j\delta(\omega-\tilde{\omega}).
\end{align}
\end{subequations}
Due to the symmetry of the indices,
we reshape $\Lambda(\omega)$ by
\begin{equation}
  \Lambda(\omega)\rightarrow\frac{1}{2}\bigg[\Lambda(\omega)+\Lambda^{T}(-\omega)\bigg],
\end{equation}
where ``$T$'' stands for matrix transposition.
Then the elements of $\Lambda(\omega)$ are modified to
\begin{equation}
\begin{aligned}
  \Lambda_{22}\rightarrow & -i\pi\left( \Delta_{ij}(\omega)+\Delta_{ij}(-\omega)\right)\\
  = & \frac{-i\omega^2}{\hbar\epsilon_0 c^2}\mathbf{d}_i\cdot
  \Im\mathbf{G}(\mathbf{r}_i, \mathbf{r}_j, |\omega|)\cdot\mathbf{d}_j \\
  = & \mathrm{sign}(\omega)\frac{-i\omega^2}{\hbar\epsilon_0 c^2}
  \mathbf{d}_i\cdot\Im\mathbf{G}(\mathbf{r}_i, \mathbf{r}_j, \omega)\cdot\mathbf{d}_j,
\end{aligned}
\end{equation}
where we have used the relation $\mathbf{G}(\omega)=\mathbf{G}^*(-\omega)$, and
\begin{equation}
  \Lambda_{21}\rightarrow  \frac{1}{2}\bigg(
   F_{ij}(\omega)+i\pi\Delta_{ij}(\omega)+F_{ij}(-\omega)-i\pi\Delta_{ij}(-\omega)\bigg).
\end{equation}
To evaluate the expressions, we shall use the Kramers-Kronig relation. For a function
$\chi(\omega)$ which
is analytic in the closed upper half-plane of $\omega$ and vanishes like $1/|\omega|$ or
faster as $|\omega|\rightarrow\infty$, and $\chi(\omega)=\chi^*(-\omega)$,
we have
\begin{equation}
  \Re\chi(\omega)=\frac{2}{\pi}\int_{0}^{\infty}
  d\omega'\mathcal{P}\frac{\omega'\Im\chi(\omega')}{\omega'^2-\omega^2}.
\end{equation}
Applying this to $\omega^2 \mathbf{G}(\omega)$,
we obtain
\begin{equation}
  F_{ij}(\omega)+F_{ij}(-\omega)=\frac{-\omega^2}{\hbar\epsilon_0
  c^2}\mathbf{d}_i\cdot\Re\mathbf{G}(\mathbf{r}_i, \mathbf{r}_j, \omega)\cdot\mathbf{d}_j,
\end{equation}
which finally gives
\begin{subequations}
\begin{align}
&  \Lambda_{21}\rightarrow \frac{-\omega^2}{2\hbar\epsilon_0 c^2}\mathbf{d}_i
\cdot\mathbf{G}(\mathbf{r}_i, \mathbf{r}_j, \omega)\cdot\mathbf{d}_j\equiv -h_{ij},\\
&  \Lambda_{12}\rightarrow \frac{-\omega^2}{2\hbar\epsilon_0
c^2}\mathbf{d}_i\cdot\mathbf{G}^*(\mathbf{r}_i, \mathbf{r}_j, \omega)\cdot\mathbf{d}_j=-h_{ij}^*.
\end{align}
\end{subequations}
Together with $\Lambda_{11}=0$, this yields the graphene-induced
emitter-emitter coupling action $S^{(p)}_{ee}$ given in Eq. (\ref{actionp})
in the main text:
\begin{subequations}
\begin{align}
\begin{split}
 & S_{ee}^{(p)}=\sum_{i,j=1}^{N}
  \int_{-\infty}^{\infty} \frac{d\omega}{2\pi}
    \begin{pmatrix}
      \phi_{i,c} & \phi_{i,q}
    \end{pmatrix}_{-\omega}\\
    &\qquad\qquad\quad \times \begin{pmatrix}
                             0 & h_{ij}^*(\omega) \\
                             h_{ij}(\omega) & 2i\Im h_{ij}(|\omega|)
                           \end{pmatrix}
                           \begin{pmatrix}
                             \phi_{j,c} \\
                             \phi_{j,q}
                           \end{pmatrix}_\omega
\end{split}\\
\begin{split}
 & h_{ij}(\omega)= \frac{\omega^2}{2\hbar\epsilon_0c^2}
  \mathbf{d}_i\cdot\mathbf{G}(\mathbf{r}_i,\mathbf{r}_j,\omega)\cdot\mathbf{d}_j.
\end{split}
\end{align}
\end{subequations}
Note that the derivation of $S_{ee}^{(p)}$ does not discard counter-rotating-wave terms
nor apply the Markov approximation, which treats the $\omega$-dependence of the spectrum as a constant.

\subsection{Spatial Disorder}\label{secC}
We define two matrices
\begin{equation}
V^1=\sigma^x=\begin{pmatrix}
      0 & 1 \\
      1 & 0
    \end{pmatrix},\quad
V^2=i\begin{pmatrix}
      0 & -1 \\
      1 & 2\mathrm{sign}(\omega)
    \end{pmatrix}.
\end{equation}
Then, $S^{(p)}_{ee}$ can be brought to the form
\begin{equation}
  S^{(p)}_{ee}=\sum_{i,j=1}^{N}\int_\omega
  \Re{h_{ij}}(\omega)v^{(1)}_{ij}(\omega)+\Im{h_{ij}}(\omega)v^{(2)}_{ij}(\omega),
\end{equation}
where
\begin{equation}
v^{(a)}_{ij}(\omega)=\begin{pmatrix}
                       \phi_{i,c} & \phi_{i,q}
                     \end{pmatrix}_{-\omega}
\cdot V^a\cdot\begin{pmatrix}
                \phi_{j,c} \\
                \phi_{j,q}
              \end{pmatrix}_{\omega}.
\end{equation}
This form will
facilitate the Gaussian averaging over the coupling strengths
$\Re{h_{ij}}(\omega), \Im h_{ij}(\omega)$. For terms with subscript
$i\neq j$, we assume a multi-component Gaussian distribution
\begin{subequations}
\begin{align}
\begin{split}
&\quad\overline{h}_{(2)}(\omega)=\int d^3\mathbf{r}_a d^3\mathbf{r}_b
p(\mathbf{r}_a,\mathbf{r}_b)h_{ab}(\omega),
\end{split}\\
\begin{split}
  & M(\omega, \omega')=\int d^3\mathbf{r}_a d^3\mathbf{r}_b p(\mathbf{r}_a,\mathbf{r}_b)\\
& \quad\times\begin{pmatrix}
    \delta\Re h_{ab}(\omega)\delta \Re h_{ab}(\omega') &
    \delta\Re h_{ab}(\omega)\delta \Im h_{ab}(\omega') \\
    \delta\Im h_{ab}(\omega)\delta \Re h_{ab}(\omega')
    &
    \delta\Im h_{ab}(\omega)\delta \Im h_{ab}(\omega')
  \end{pmatrix}.
\end{split}
\end{align}
\end{subequations}
These are Eqs. (7a) and (7b) of the main text. Different from the
emitter-emitter coupling strength, the
values of the graphene-induced individual terms, $\Re{h_{ii}}(\omega),\, \Im h_{ii}(\omega)$,
depend only on the distance from the emitter to the graphene.
Since we have assumed that the layer of emitters is parallel to the graphene monolayer,
all the $h_{ii}(\omega)$ are fixed and identical.
In Sec. \ref{secF}, we present figures showing these coupling strengths
and the elements of the covariance matrix.

To explore the phase transition at $N\rightarrow\infty$, we define
\begin{equation}\label{scale}
  h^{d}_i=N\times{h}_{ii},\quad h^{o}=N\times \bar{h}_{(2)},
  \quad M^{o}=N\times M,
\end{equation}
so that after averaging over $h_{ij} (i\neq j)$ as described in the main text, we have
\begin{equation}\label{barS}
\begin{aligned}
  \bar{S}^{(p)}_{ee}=& \frac{1}{N}\sum_{i=1}^{N} \int_\omega
  (h^d_i-h^o)_{a}(\omega)v^{(a)}_{ii}(\omega)\\
  & +\frac{1}{N}\sum_{i,j=1}^N\int_\omega
   h^o_{a}(\omega)v^{(a)}_{ij}(\omega) \\
  & +i\frac{1}{N}\sum_{i\neq j=1}^N \int_{\omega,\omega'}v^{(a)}_{ij}(\omega)
  M^o_{ab}(\omega,\omega')v^{(b)}_{ij}(\omega'),
\end{aligned}
\end{equation}
where the summation over replicated indices $a,b$ are implicit assumed; and
we have written $h^{d(o)}$ in the vector form of $(\Re h^{d(o)}, \Im h^{d(o)})$.
While, in the third line of Eq. (\ref{barS}), terms with $i=j$ are excluded, in the limit of large $N$, we may
release this exclusion (see more discussion in Sec. \ref{IB}) and define
\begin{equation}
\begin{aligned}
  \Phi_\alpha(\omega)=& \sum_{i=1}^N\phi_{i,\alpha}(\omega), \\
   \Phi_{\alpha\beta}(\omega,\omega')=&
   \sum_{i=1}^{N}\phi_{i,\alpha}(\omega)\phi_{i,\beta}(\omega').
\end{aligned}
\end{equation}
Now the Keldysh action can be expressed in terms of $\Phi_\alpha$ and
$\Phi_{\alpha\beta}$:
\begin{equation}\label{S-res}
\begin{aligned}
  S= & \frac{1}{N}\sum_{i=1}^{N}\int_{\omega}\phi_{i,\alpha}(-\omega)\phi_{i,\beta}(\omega)
  \Lambda^{e}_{i,\alpha\beta}(\omega)-2\sum_{i=1}^{N}\lambda_{i,q}2\pi\delta(0) \\
  & +\frac{1}{N}\int_\omega \Phi_\alpha(-\omega)\Phi_\beta(\omega)\Lambda^{ce}_{\alpha\beta}(\omega) \\
  & +i\frac{1}{N}\int_{\omega,\omega'}\Phi_{\alpha\beta}(-\omega,-\omega')
  \tilde{M}_{\alpha\beta,\alpha'\beta'}(\omega,\omega')
  \Phi_{\alpha'\beta'}(\omega,\omega'),
\end{aligned}
\end{equation}
where the new matrixes are defined as
\begin{equation}\label{tilM}
\begin{aligned}
&  \Lambda^{e}_i =N\begin{pmatrix}
                   \lambda_{i,q} & \lambda_{i,c}-\frac{\omega^2}{\omega_i} \\
                   \lambda_{i,c}-\frac{\omega^2}{\omega_i} & \lambda_{i,q}
                 \end{pmatrix}+(h^d_i-h^o)_a V^a, \\
& \Lambda^{ce} = h^o_a V^a, \\
& \tilde{M}_{\alpha\beta, \alpha'\beta'}(\omega,\omega') =
\sum_{s,t}V^s_{\alpha\alpha'}(\omega)M^o_{st}(\omega,\omega')
V^t_{\beta\beta'}(\omega').
\end{aligned}
\end{equation}

Then we apply the Hubbard-Stratonovich transformation \cite{Stratonovich1957,Hubbard1959}
based on the formula that
\begin{equation}
\begin{aligned}
& \int D[\psi_\alpha]
e^{-iN\int_{\omega}\psi_{\alpha}(-\omega)\Lambda_{\alpha\beta}^{ce}(\omega)\psi_{\beta}(\omega)
-2i\int_{\omega}\psi_{\alpha}(-\omega)\Lambda_{\alpha\beta}^{ce}(\omega)\phi_{\beta}(\omega)}\\
&\propto e^{i\frac{1}{N}\int_{\omega}\phi_{\alpha}(-\omega)
\Lambda_{\alpha\beta}^{ce}(\omega)\phi_{\beta}(\omega)},
\end{aligned}
\end{equation}
The coefficient of proportionality in the above formula is a constant,
which is irrelevant to the dynamical variables.
The Hubbard-Stratonovich transformation of $\Phi_{\alpha\beta}$
is based on a similar formula of Gaussian integral
\begin{equation}\label{STm}
\begin{aligned}
& \int D[Q_{a}]e^{-N\int_{q}Q_{a}(-q)\tilde{M}_{ab}(q)Q_{b}(q)
-2i\int_{q}Q_{\alpha}(-q)\tilde{M}_{ab}(q)\Phi_{b}(q)}\\
& \propto e^{-\frac{1}{N}\int_{q}\Phi_{a}(-q)\tilde{M}_{ab}(q)\Phi_{b}(q)},
\end{aligned}
\end{equation}
where ``a'' and ``b'' denote the subscript $(\alpha\beta)$ and
$(\alpha'\beta')$,
and ``q'' is used to abbreviate $(\omega,\omega')$.

After the transformations,
the Keldysh action has some residual $\phi_i$ terms
of order less than or equal to two.
We can eliminate these terms by Gaussian integrals.

Then the Keldysh action becomes a functional of
the Lagrange multiplier $\lambda_{i,\alpha}$ and the
two new dynamical variables,
$\psi_\alpha$ and $Q_{\alpha\beta}$, which are introduced
in Eq. (\ref{Q}) of the main text. Substituting
the static ansatz at mean field level,
\begin{equation}
\begin{aligned}
   \psi_\alpha(\omega)& =2\pi\psi_\alpha\delta(0), \\
   Q_{\alpha\beta}(\omega,\omega')& =Q_{\alpha\beta}(\omega)2\pi\delta(\omega+\omega'),
\end{aligned}
\end{equation}
this finally yields the action in terms of $\psi_\alpha,
Q_{\alpha\beta}$ and $\lambda_i$
\begin{equation}\label{S-sum}
 \begin{aligned}
  S= & \frac{i}{2}\sum_{i=1}^{N}\mathrm{tr}\ln(2\mathbf{L}_i)
   -2\sum_{i=1}^{N}\pi\delta(0)(\Lambda^{ce}\psi)^T\mathbf{L}_i^{-1}(0)
   (\Lambda^{ce}\psi)\\
   & +i2\pi\delta(0)N\int_{\omega}Q_{\alpha\beta}(-\omega)
   \tilde{M}_{\alpha\beta,\alpha'\beta'}(\omega,-\omega)Q_{\alpha'\beta'}(\omega)\\
   & -2\pi\delta(0)N\psi_\alpha\Lambda^{ce}_{\alpha\alpha'}(0)\psi_{\alpha'}
   -4\pi\delta(0)\sum_{i=1}^{N}\lambda_{i,q}.
\end{aligned}
\end{equation}
where the matrix $\mathbf{L}_i$ is defined as
\begin{equation}\label{L}
\begin{aligned}
  &\mathbf{L}_i(\omega,\omega') = \mathbf{L}(\omega)2\pi\delta(\omega+\omega') \\
  &\mathbf{L}_{i,\alpha\beta}(\omega) =\frac{1}{N}\Lambda^{e}_{i,\alpha\beta}(-\omega)
  -2Q_{\alpha'\beta'}(-\omega)\tilde{M}_{\alpha'\beta',\alpha\beta}(\omega,-\omega).
\end{aligned}
\end{equation}

\subsection{Saddle Point Equations}\label{secD}
We now turn to the solution of the saddle point equations
\begin{equation}
  \frac{\delta}{\delta q} S \stackrel{!}{=}0, \quad q\in\{\lambda_{i,\alpha},
  \psi_\alpha, Q_{\alpha\beta}\},
\end{equation}
which is restricted by the causality conditions $\lambda_q= Q_{qq}=\psi_q=0$.

\subsubsection{Equations for $\lambda_{i,\alpha}$}
We assume $\lambda_{i,\alpha}=\lambda_\alpha$,
and replace the summation in Eq. (\ref{S-sum}) with a factor of N.
The saddle point equation with respect to the
Lagrangian multiplier $\lambda_q$ is
\begin{equation}\label{eq-lambda}
\begin{aligned}
  \frac{i}{2}\int_\omega\mathrm{tr} [ & \mathbf{L}^{-1}_{reg}(\omega)] -
  \frac{1}{\det \mathbf{L}(0)} \bigg(\psi_c^2(\Lambda^{ce}_{cq})^2\\
  + & q_{EA}
  \tilde{M}_{cc,qq}(0)\bigg)-2=0,
\end{aligned}
\end{equation}
which confirms the restriction $\phi_i^2=1$.
In Eq. (\ref{eq-lambda}), $\mathbf{L}_{reg}$ refers to the part defined with $Q_{cc}^{reg}$.
The equation with respect to the Lagrangian multiplier $\lambda_c$ is
\begin{equation}
  -\frac{i}{2}\int_\omega \frac{1}{\det \mathbf{L}(\omega)}\mathrm{tr}[\sigma_x \mathbf{L}(\omega)]=0.
\end{equation}
This equation is a statement of the universal property of the Keldysh Green's function that
\begin{equation}
  Q^R(t,t)+Q^{A}(t,t)=0,
\end{equation}
where $Q^R=Q_{cq}$ and $Q^A=Q_{qc}$.

\subsubsection{Equations for $\psi_\alpha$}
For $\psi_c$, the saddle-point equation is trivial,
because
\begin{equation}
(\Lambda^{ce})_{cc}=0, (\Lambda^{ce}\mathbf{L}^{-1}\Lambda^{ce})_{cc}=0,
 (\mathbf{L}^{-1})_{qq}=0,
\end{equation}
when $\lambda_q=\psi_q=Q_{qq}=0$.

For $\psi_q$, the saddle-point equation gives
\begin{equation}\label{psii-c}
  \psi_c\big( \Lambda^{ce}_{qc}(\mathbf{L}^{-1})_{cq}+1 \big)=0,
\end{equation}
which gauges the relation between $\lambda_c$ and $Q_{cq}$, in the
SR phase where $\psi_c\neq 0$.

\subsubsection{Equations for $Q^{reg}_{\alpha\beta}$}

The Edwards-Anderson order parameter $q_{EA}$ is introduced as the singular
part of $Q_{cc}(\omega)$:
\begin{equation}
  Q_{cc}(\omega)=Q^{reg}_{cc}(\omega)-2\pi iq_{EA}\delta(\omega),
\end{equation}
and the saddle point equation for the regular component reads
\begin{equation}
\begin{aligned}
  2Q^{reg}_{\alpha\beta}(\omega)= &
  [\mathbf{L}(\omega)]^{-1}_{reg,\beta\alpha} + 4i\pi\bigg(\frac{\psi_c^2(\Lambda^{ce}_{cq})^2}
  {\det \mathbf{L}(0)}\\
  & +q_{EA}+q_{EA}\frac{\tilde{M}_{cc,qq}(0)}
  {\det\mathbf{L}(0)}\bigg)\delta_{\alpha c}\delta_{\beta c}\delta(\omega).
\end{aligned}
\end{equation}
Note that, this equation can be separated into the
regular part and the singular part at $\omega=0$:
\begin{equation}
\begin{aligned}
  2Q^{reg}_{\alpha\beta} &=[\mathbf{L}(\omega)]^{-1}_{reg,\beta\alpha};\\
  (\Lambda^{ce}_{cq})^2\psi_c^2 &=-q_{EA}(\tilde{M}_{cc,qq}+\det\mathbf{L}(0)).
\end{aligned}\label{qqab}
\end{equation}
For the regular part, implementing the substitution of
Eq. (\ref{L}) for the $cq$ component gives
\begin{equation}\label{qcq}
  \frac{1}{2Q_{cq}}=\lambda_c-\frac{\omega^2}{\omega_z}+\bar{h}_{(1)}
-\bar{h}_{(2)}-2Q_{cq}\tilde{M}_{qc,cq}(\omega,-\omega).
\end{equation}
where we have assumed $\lambda_{i,\alpha}=\lambda_\alpha$ for every
emitter.

We find that this equation does not have a unique solution except
in the absence of randomness, $M\rightarrow 0$, where
 the second line of Eq. (\ref{S-sum}) vanishes
and the Keldysh action attains the value given in Ref. \cite{Torre2013}.
We select the solution that is
continuously connected to the unique solution to Eq. (\ref{qcq}) with $\tilde{M}=0$, under the variation of $\lambda\tilde{M}, \lambda:1\rightarrow 0$.

The regular part of $Q_{cc}$ turns out to be
\begin{equation}
\begin{aligned}
 & Q_{cc}^{reg}= \frac{4|Q_{cq}|^2}{1-4|Q_{cq}|^2
  \tilde{M}_{cc,qq}(\omega,-\omega)}\bigg( Q_{qc}\tilde{M}_{cq,qq}(\omega,-\omega)\\
  & +Q_{cq}\tilde{M}_{qc,qq}(\omega,-\omega)-i\mathrm{sgn}(\omega)
  (\Im\bar{h}_{(1)}-\Im\bar{h}_{(2)})\bigg).
  \end{aligned}
\end{equation}
The causality condition of the Keldysh formalism implies
$Q_{qq}=\lambda_{i,q}=\psi_q=0$,
and $Q_{cq}(\omega)=Q_{qc}^*(\omega)$ \cite{Sieberer2016}.

Since the Edward-Anderson order parameter $q_{EA}$ is non-negative,
it follows from the second equation of Eq. (\ref{qqab})
that to have $\psi_c^2>0$, we must have
\begin{equation}
  \tilde{M}_{cc,qq}(0,0)+\det\mathbf{L}(0)<0.
\end{equation}
This relation helps to distinguish the SR phase and the SG phase.

\subsubsection{Determination of $\lambda_c$ and the three phases}
In the SR phase, $\psi_c\neq 0$, so that
Eq. (\ref{psii-c}) determines the value of $\lambda_c^{SR}$:
\begin{equation}
  \lambda_{c}^{SR}=-\bar{h}_{(1)}(0)+\bar{h}_{(2)}(0)-\Lambda_{qc}-
  \frac{N}{\Lambda_{qc}}M_{11}(0,0),
\end{equation}
where $M_{11}$ is the real-real element of $M$, and
$\Lambda_{qc}=N\Re\overline{h}_{(2)}(0)$.

In the SG phase, we have $q_{EA}>0$ and $\psi_c=0$.
Therefore, the singular part of Eq. (\ref{qqab}) yields
\begin{equation}
\tilde{M}_{cc,qq}(0,0)+\det\mathbf{L}(0)=0.
\end{equation}
Note that $\det\mathbf{L}(0)=\frac{1}{4Q_{cq}Q_{qc}(0)}$.
Corresponding to cases $\frac{1}{2Q_{cq}(0)}=\pm \sqrt{\tilde{M}_{qc,cq}(0,0)}$,
we have
\begin{equation}
  \lambda_c^{SG}=-\frac{1}{N}
  \bigg(h^d(0)-h^o(0)\bigg)\pm 2\sqrt{N\times M_{11}(0)}.
\end{equation}
The possibility of $\lambda_c^{SR}=\lambda_c^{SG}$ corresponds to the
minus sign of the above equation. Thus we get
\begin{equation}
  \lambda_c^{SG}=-\bar{h}_{(1)}(0)+\bar{h}_{(2)}(0)-2\sqrt{N\times M_{11}(0,0)}.
\end{equation}

It turns out that
the system is in the SR phase rather than the SG phase only if
\begin{equation}
  \bigg(\overline{h}_{(2)}(0)\bigg)^2
  >\frac{1}{N}M_{11}(0,0).
\end{equation}
This expression also gives the analytical result of the SG-SR phase
boundary. In Sec. \ref{secF} we will elaborate on the calculation for
the emitter-graphene system. We find that the values of $\overline{h}_{(2)}(0)$
and $M_{11}(0,0)$ are insensitive to the graphene Fermi energy $E_f$.

For the normal phase, $\lambda_c$ should be determined from the equality
\begin{equation}
\frac{i}{4\pi}\int_{-\infty}^{\infty}d\omega Q^{reg}_{cc}(\omega)=2.
\end{equation}
The boundaries between the normal phase and the other phases are obtained
by matching their values of $\lambda_c$.

The determination of $q_{EA}$ and $\psi_c$, which are present in the
singular part of $Q_{cc}(\omega)$, are obtained from the equality
\begin{equation}
\frac{i}{4\pi}\int_{-\infty}^{\infty}d\omega Q_{cc}(\omega)=2.
\end{equation}

\subsection{Inhomogeneous Broadening}\label{IB}
We suppose the emitters suffer from inhomogeneous broadening
so that the transition frequency follows a Gaussian distribution
\begin{equation}
  \rho(\omega_{i,z})=\frac{1}{\sqrt{2\pi}\Delta}\exp\bigg(
  -\frac{(\omega_{i,z}-\omega_z)^2}{2\Delta^2}\bigg),
\end{equation}
where $\Delta$ is the standard deviation of $\omega_{i,z}$.
The corresponding probability distribution of $\frac{1}{\omega_{i,z}}$, is
\begin{equation}
\begin{aligned}
p(\frac{1}{\omega_{i,z}})=& \omega_{i,z}^2\rho(\omega_{i,z})\\
=&\frac{1}{\sqrt{2\pi}\Delta}\exp\bigg(2\ln \omega_{i,z}
  -\frac{(\omega_{i,z}-\omega_z)^2}{2\Delta^2}\bigg).
\end{aligned}
\end{equation}
The condition $\Delta\ll \omega_z$ implies
that $\ln \omega_{i,z}\approx \ln\omega_z+\omega_{i,z}/\omega_z-1$.
Thus, $1/\omega_{i,z}$
has a Gaussian distribution with variance $\frac{\Delta}{\omega^2_z}$,
\begin{equation}
  p(\frac{1}{\omega_{i,z}})\approx \frac{\omega_z^2}{\sqrt{2\pi}\Delta}
  \exp\bigg( -\frac{(1/\omega_{i,z}-1/\omega_z)^2}{2(\Delta/\omega_z^2)^2}\bigg).
\end{equation}
We shall average functions of  $\omega_{i,z}$ according to this distribution.
Let us rewrite the Keldysh action of the free emitters, Eq. (\ref{action-e}) of the
main text, but replace $\omega_z$ with $\omega_{i,z}$:
\begin{equation}
  S_e=-\sum_{i=1}^{N}\sum_{a=\pm}a\int dt\frac{1}{\omega_{i,z}}(\partial_t \phi_{i,a})^2+
  \lambda_{i,a}(t)(\phi_{i,a}^2-1).
\end{equation}
Compared with the Keldysh action without inhomogeneous
broadening, an additional
term is obtained from the average of $\omega_{i,z}$, that is,
\begin{equation}
\begin{aligned}
  S^{(b)}=& i\frac{\Delta^2}{2\omega_z^4}\sum_{i=1}^{N}\int_{\omega,\omega'}
     \omega^2\omega'^2\\
     & \qquad \times \phi_{i,c}(-\omega)\phi_{i,q}(\omega)
     \phi_{i,c}(-\omega')\phi_{i,q}(\omega').
\end{aligned}
\end{equation}
Note that the integrals over $\omega$ and $\omega'$ are independent and factor into a product.
We recall that in Eq. (\ref{barS}) we made an approximation and released the restriction that $i\neq j$. We can reintroduce
the restriction by
incorporating the individual terms with $i=j$, and obtain the action
\begin{equation}\label{S-add}
  S^{(b)}-i\frac{1}{N}\sum_{i=1}^{N}\int_{\omega,\omega'}
  v^{(a)}_{ii}(\omega)M^o_{ab}(\omega,\omega')v^{(b)}_{ii}(\omega').
\end{equation}
To cope with the 4-order terms, we shall apply the
Hubbard-Stratonovich transformation.

Let us define $\Phi^i_{\alpha\beta}(\omega,\omega')=\phi_{i,\alpha}(\omega)\phi_{i,\beta}(\omega')$.
Then Eq. (\ref{S-add}) can be rewritten as
\begin{equation}\label{newm}
  i\sum_{i=1}^{N}\int_{\omega,\omega'}\Phi^i_{\alpha\beta}(-\omega,-\omega')
  \delta\tilde{M}_{\alpha\beta,\alpha'\beta'}(\omega,\omega')
  \Phi^i_{\alpha'\beta'}(\omega,\omega').
\end{equation}
where the matrix $\delta\tilde{M}_{\alpha\beta,\alpha'\beta'}(\omega,\omega')$
is defined as
\begin{equation}
\begin{aligned}\label{eff-broad}
  & \delta\tilde{M}_{\alpha\beta,\alpha'\beta'}(\omega,\omega')=
  -\frac{1}{N}\tilde{M}_{\alpha\beta,\alpha'\beta'}(\omega,\omega')\\
  &\qquad\quad +\omega^2\omega'^2\frac{\Delta^2}{8\omega_z^4}
 (\delta_{\alpha\beta, cq}
  \delta_{\alpha'\beta',qc}+\delta_{\alpha\beta,qc}\delta_{\alpha'\beta',cq}\\
  & \qquad\quad \qquad +\delta_{\alpha\beta,cc}\delta_{\alpha'\beta',qq}
   + \delta_{\alpha\beta,qq}\delta_{\alpha'\beta',cc}).
\end{aligned}
\end{equation}
We can implement the Hubbard-Stratonovich transformation of Eq. (\ref{newm})
in a way similar to Eq. (\ref{STm}):
\begin{equation}
\begin{aligned}
& \int D[Q^i_{a}]e^{-N\int_{q}Q^i_{a}(-q)\tilde{M}_{ab}(q)Q^i_{b}(q)
-2i\int_{q}Q^i_{\alpha}(-q)\tilde{M}_{ab}(q)\Phi_{b}(q)}\\
& \propto e^{-\frac{1}{N}\int_{q}\Phi^i_{a}(-q)\tilde{M}_{ab}(q)\Phi^i_{b}(q)},
\end{aligned}
\end{equation}
where the conventions of notation are the same as in Eq. (\ref{STm}).
In the sense of saddle-point equations, the physical meaning
of $Q^i_{\alpha\beta}$ is
\begin{equation}
  Q^i_{\alpha\beta}=\langle \phi_{i,\alpha}\phi_{i,\beta}\rangle
\end{equation}
By the further assumption of the homogeneous mean-field ansatz,
that for $\forall i$,
\begin{equation}
  \langle \phi_{i,\alpha}\phi_{i,\beta}\rangle=
  \frac{1}{N}\sum_{k=1}^{N}\langle\phi_{k,\alpha}\phi_{k,\beta}\rangle,
\end{equation}
we can replace the new variable $Q^i_{\alpha\beta}$ with $Q_{\alpha\beta}$,
which is defined in the context of spatial disorders.

The result of all the above steps can also be obtained by rewriting Eq. (\ref{S-add}) as
\begin{equation}
  i\frac{1}{N}\int_{\omega,\omega'}\Phi_{\alpha\beta}(-\omega,-\omega')
  \delta\tilde{M}_{\alpha\beta,\alpha'\beta'}(\omega,\omega')
  \Phi_{\alpha'\beta'}(\omega,\omega')
\end{equation}
followed by the Hubbard-Stratonovich transformation in a way similar to Eq. (\ref{STm}).
It means that, the effect of inhomogeneous broadening
can be seen as a modification of the matrix $\tilde{M}$
defined in Eq. (\ref{S-res}) by a term $\delta\tilde{M}$
given in Eq. (\ref{newm}).

Note that the first term of Eq. (\ref{newm}) comes from the additional term
mentioned in Eq. (\ref{S-add}) and contributes only little when $N\gg 1$,
thus justifying the approximation made for Eq. (\ref{S-res}).
Since $\tilde{M}$ is defined with a
factor of $N$, see Eq. (\ref{scale}), the correction
made by inhomogeneous broadening, Eq. (\ref{eff-broad}),
is also negligible when $N$ is large.

\subsection{The specific example of the Emitter-Graphene System}\label{secF}
The surface conductivity of the graphene monolayer is
\begin{equation}
  \begin{aligned}
  \sigma(E_f,& \tau; \omega)= \frac{e^2E_f}{\pi\hbar^2}\frac{i}{\omega+i\tau^{-1}}\\
  & +\frac{e^2}{4\hbar}\bigg( \Theta(\hbar\omega-2E_f)+\frac{i}{\pi}\log
  |\frac{\hbar\omega-2E_f}{\hbar\omega+2E_f}|\bigg).
  \end{aligned}
\end{equation}

When the emitter dipoles are aligned perpendicular to the graphene monolayer,
the relevant element of the dyadic Green's tensor is
$\mathbf{G}^0_{zz}+\mathbf{G}^s_{zz}$, where $\mathbf{G}^0_{zz}$ is the
vacuum dyadic Green's function for
free propagation modes and
$\mathbf{G}^s_{zz}$ is the so-called `scattering' part accounting for
the surface plasmon modes of the graphene monolayer
\begin{equation}\label{gzzf}
\begin{aligned}
  \frac{\omega^2}{c^2}\mathbf{G}^s_{zz}& (r,r';z)=
  \int \frac{d^2\mathbf{k}_\shortparallel}{(2\pi)^2}\frac{i}{2\epsilon_1 k_{1,z}}
  k^2_\shortparallel r_p e^{i\mathbf{k}_\shortparallel\cdot\mathbf{\delta r}+2ik_{1,z}z}\\
  &=\int \frac{dk_\shortparallel}{2\pi}\frac{i}{2\epsilon_1 k_{1,z}}
  k_\shortparallel^3 r_p J_0(k_\shortparallel \delta r),
  e^{2ik_{1,z}z}
\end{aligned}
\end{equation}
where $\mathbf{\delta r}=r-r'$ and $\delta r$ is its length;
$J_0$ is the zero-order Bessel function;
$\epsilon_{1(2)}$ is the relative permittivity of the
dielectric above(below) the
graphene monolayer, $r_p$ is the Fresnel coefficient of reflection of the p-modes from above the graphene layer
\begin{equation}
  r_p=\frac{-\epsilon_1 k_{2,z}+\epsilon_{2} k_{1,z}+
  \frac{\sigma(\omega)}{\omega\epsilon_0}k_{1,z}k_{2,z}}
  {\epsilon_1 k_{2,z}+\epsilon_{2} k_{1,z}+
  \frac{\sigma(\omega)}{\omega\epsilon_0}k_{1,z}k_{2,z}},
\end{equation}
where $k_{1(2),z}=\sqrt{\frac{\omega^2}{c^2}\epsilon_{1(2)}-k_\shortparallel^2}$.
Note that in the limit $\omega\rightarrow 0$, $r_p$ equals 1 and
does not depend on the Fermi energy. As a result,
the Fermi energy $E_f$ is irrelevant to the SG-SR boundary.

In the numerical calculation, it is convenient to normalize
$k_\shortparallel$ and $\delta r$ in the above expressions by
$\omega_z/c$. That is, define
\begin{equation}
  k_\shortparallel=\frac{\omega}{c}\tilde{k}_\shortparallel,
  \quad
  \delta r=\frac{c}{\omega_z}\delta\tilde{r},
  \quad
  z=\frac{c}{\omega_z}\tilde{z},
\end{equation}
and then Eq. (\ref{gzzf}) is recast to
\begin{equation}
  (\frac{\omega_z}{c})^3\int_{0}^{\infty}\frac{d\tilde{k}_\shortparallel}{2\pi}
  \frac{i}{2\epsilon_1\tilde{k}_{1,z}}\tilde{k}_\shortparallel^3 r_p
  J_0(\tilde{k}_\shortparallel\delta\tilde{r})e^{2i\tilde{k}_{1,z}\tilde{z}}.
\end{equation}
The factor $(\frac{\omega_z}{c})^3$ can then be combined with the
length of $\mathbf{d}_i$ and absorbed into the expression for the
vacuum spontaneous emission rate $\gamma_0$.

\begin{figure}[tb]\label{sp-figure}
  \centering
  \includegraphics[width=\textwidth]{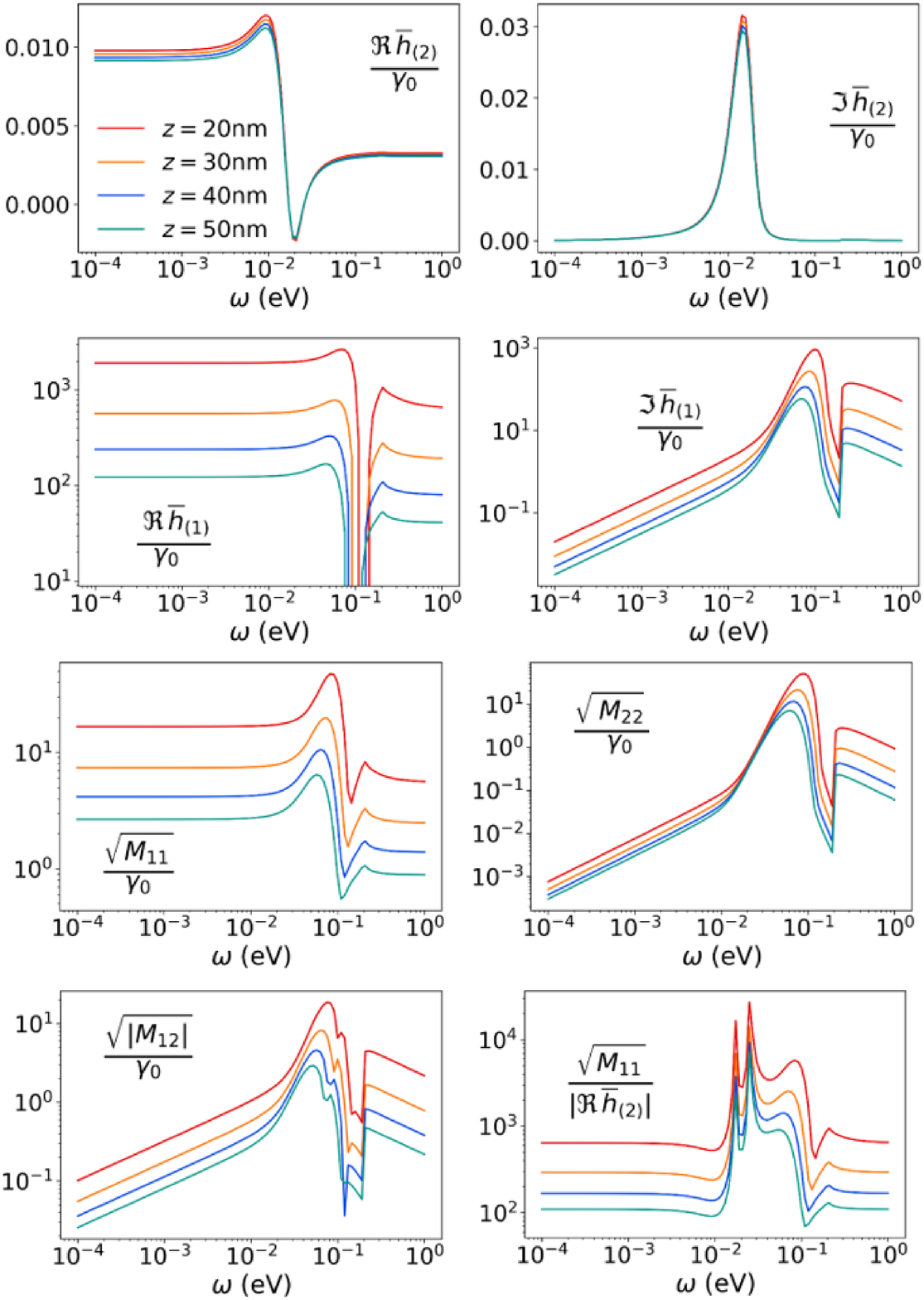}
  \caption{Coefficients of the emitter-graphene surface plasmon coupling,
  for systems with $\omega_z=0.5\,\mathrm{eV}, E_f= 0.1\,\mathrm{eV}$, and $L=10^3$ nm.
  From top to the bottom in the figures we show results for the different heights $z$=20 (red),
  30 (orange), 40 (blue) and 50 (green) nm. The dimensionless values are normalized by the
  emitter spontaneous emission rate $\gamma_0$.}
\end{figure}

The surface-plasmons have the dispersion relation
\begin{equation}
\epsilon_1 k^{sp}_{2,z}+\epsilon_2 k^{sp}_{1,z} +
\frac{\sigma(\omega_{sp})}{\omega_{sp}\epsilon_0}k^{sp}_{1,z}k^{sp}_{2,z}=0,
\end{equation}
where $k_{1(2),z}=\sqrt{\frac{\omega^2}{c^2}\epsilon_{1(2)}-k_{sp}^2}$,
$\omega_{sp}$ and $k_{sp}$ represent the frequency
and wavevector of the surface-plasmon, respectively.

The horizontal coordinates $\{(x_i, y_i)\}_i$ of the
emitters are assumed
to follow the identical Gaussian distribution
\begin{equation}
  p(x,y)=\frac{1}{2\pi L^2}\exp(-\frac{x^2+y^2}{2L^2}),
\end{equation}
and the distance between any two emitters follows the distribution
\begin{equation}
  p_L(\delta r)=\frac{\delta r}{2 L^2}\exp (-\frac{(\delta r)^2}{4 L^2}).
\end{equation}
To calculate the mean values and covariances
$\overline{h}_{(1)}(\omega), \overline{h}_{(2)}(\omega)$ and
$\tilde{M}(\omega,\omega')$ required in our formalism,
the use of the Gaussian distribution permits analytical handling of the oscillating integrants
related to the Bessel function $J_0(k\delta r)$.
\begin{equation}
\begin{aligned}
  &\int_0^\infty dr\; \frac{r}{2L^2}J_0(kr)\exp(\frac{-r^2}{4L^2})=\exp(-k^2L^2),\\
  &\int_{0}^{\infty}dr\;\frac{r}{2L^2}J_0(kr)J_0(k'r)\exp(-\frac{r^2}{4L^2})=\\
  &\qquad\qquad I_0(2L^2 k k')\exp\bigg(-L^2(k^2+k'^2)\bigg).
\end{aligned}
\end{equation}
where $I_0$ is the modified Bessel function.
By use of these formulas, the remaining integrals are numerically well behaved.

Finally, to have an impression of the numerical results, we illustrate
the $z$-dependence of the averaged coupling strength and the elements
of the covariance matrix in Fig. 4. It shows that by decreasing $z$, the graphene SP-induced
self-interaction terms and the elements of the covariance matrix are increased significantly,
while the SP-induced emitter-emitter coupling strength changes little. It confirms
our argument about the $z$-dependence made in the main text.


\begin{thebibliography}{70}%
\bibitem{Dicke1954} R. H. Dicke, Phys. Rev. \textbf{93}, 99 (1954).
\bibitem{Hepp1973} K. Hepp and E. H. Lieb, Annals of Physics \textbf{76}, 360
(1973).
\bibitem{Wang1973} Y. K. Wang and F. T. Hioe, Phys. Rev. A \textbf{7}, 831 (1973).
\bibitem{Emeljanov1976} V. Emeljanov and Y. Klimontovich, Physics Letters A
\textbf{59}, 366 (1976).
\bibitem{nogo1981} K. Gaw\c{e}dzki and K. Rza\c{\'{z}}ewski, Phys. Rev. A \textbf{23}, 2134
(1981).
\bibitem{Bamba2014} M. Bamba and T. Ogawa, Phys. Rev. A \textbf{90}, 063825
(2014).
\bibitem{Keeling2007} J. Keeling, Journal of Physics: Condensed Matter \textbf{19},
295213 (2007).
\bibitem{Vukics2012} A. Vukics and P. Domokos, Phys. Rev. A \textbf{86}, 053807
(2012).
\bibitem{Vukics2014} A. Vukics, T. Grie{\ss}er, and P. Domokos, Phys. Rev. Lett.
\textbf{112}, 073601 (2014).
\bibitem{Vukics2015} A. Vukics, T. Grie{\ss}er, and P. Domokos, Phys. Rev. A
\textbf{92}, 043835 (2015).
\bibitem{Grieser2016} T. Grie{\ss}er, A. Vukics, and P. Domokos, Phys. Rev. A
\textbf{94}, 033815 (2016).
\bibitem{Baumann2010} K. Baumann, C. Guerlin, F. Brennecke, and
T. Esslinger, Nature \textbf{464}, 1301 (2010).
\bibitem{Baumann2011} K. Baumann, R. Mottl, F. Brennecke, and T. Esslinger,
Phys. Rev. Lett. \textbf{107}, 140402 (2011).
\bibitem{Brennecke2013} F. Brennecke, R. Mottl, K. Baumann, R. Landig, T. Donner, and T. Esslinger, Proceedings of the National
Academy of Sciences \textbf{110}, 11763 (2013).
\bibitem{Baden2014} M. P. Baden, K. J. Arnold, A. L. Grimsmo, S. Parkins,
and M. D. Barrett, Phys. Rev. Lett. \textbf{113}, 020408 (2014).
\bibitem{Klinder2015} J. Klinder, H. Ke{\ss}ler, M. Wolke, L. Mathey, and A. Hemmerich, Proceedings of the National Academy of Sciences
\textbf{112}, 3290 (2015).
\bibitem{Zhiqiang2017} Z. Zhiqiang, C. H. Lee, R. Kumar, K. J. Arnold, S. J.
Masson, A. S. Parkins, and M. D. Barrett, Optica 4,
\textbf{424} (2017).
\bibitem{Dimer2007} F. Dimer, B. Estienne, A. S. Parkins, and H. J.
Carmichael, Phys. Rev. A \textbf{75}, 013804 (2007).
\bibitem{Tolkunov2007} D. Tolkunov and D. Solenov, Phys. Rev. B \textbf{75}, 024402
(2007).
\bibitem{Gopalakrishnan2011} S. Gopalakrishnan, B. L. Lev, and P. M. Goldbart, Phys.
Rev. Lett. \textbf{107}, 277201 (2011).
\bibitem{Strack2011} P. Strack and S. Sachdev, Phys. Rev. Lett. \textbf{107}, 277202
(2011).
\bibitem{Sieberer2016} L. M. Sieberer, M. Buchhold, and S. Diehl, Reports on
Progress in Physics \textbf{79}, 096001 (2016).
\bibitem{Buchhold2013} M. Buchhold, P. Strack, S. Sachdev, and S. Diehl, Phys.
Rev. A \textbf{87}, 063622 (2013).
\bibitem{Torre2013} E. G. D. Torre, S. Diehl, M. D. Lukin, S. Sachdev, and
P. Strack, Phys. Rev. A \textbf{87}, 023831 (2013).
\bibitem{Kirton2017} P. Kirton and J. Keeling, Phys. Rev. Lett. \textbf{118}, 123602
(2017).
\bibitem{Bastidas2012} V. M. Bastidas, C. Emary, B. Regler, and T. Brandes,
Phys. Rev. Lett. \textbf{108}, 043003 (2012).
\bibitem{Viehmann2011} O. Viehmann, J. von Delft, and F. Marquardt, Phys.
Rev. Lett. \textbf{107}, 113602 (2011)
\bibitem{Bamba2016} M. Bamba, K. Inomata, and Y. Nakamura, Phys. Rev.
Lett. \textbf{117}, 173601 (2016).
\bibitem{Zou2014} L. J. Zou, D. Marcos, S. Diehl, S. Putz, J. Schmiedmayer,
J. Majer, and P. Rabl, Phys. Rev. Lett. \textbf{113}, 023603
(2014).
\bibitem{Lambert2004} N. Lambert, C. Emary, and T. Brandes, Phys. Rev. Lett.
\textbf{92}, 073602 (2004).
\bibitem{Emary2003} C. Emary and T. Brandes, Phys. Rev. Lett. \textbf{90}, 044101
(2003).
\bibitem{Pustovit2009} V. N. Pustovit and T. V. Shahbazyan, Phys. Rev. Lett.
\textbf{102}, 077401 (2009).
\bibitem{Tame2013} M. S. Tame, K. R. McEnery, . K. \"{O}zdemir, J. Lee, S. A.
Maier, and M. S. Kim, Nature Physics \textbf{9}, 329 (2013).
\bibitem{Toermae2015} P. T\"{o}rm\"{a} and W. L. Barnes, Reports on Progress in Physics
\textbf{78}, 013901 (2015).
\bibitem{Chang2006} D. E. Chang, A. S. S{\o}rensen, P. R. Hemmer, and M. D.
Lukin, Phys. Rev. Lett. \textbf{97}, 053002 (2006).
\bibitem{Zhang2006} W. Zhang, A. O. Govorov, and G. W. Bryant, Phys.
Rev. Lett. \textbf{97}, 146804 (2006).
\bibitem{Basov2016} D. N. Basov, M. M. Fogler, and F. J. Garc\'{\i}a de Abajo,
Science \textbf{354} (2016).
\bibitem{Grigorenko2012} A. N. Grigorenko, M. Polini, and K. S. Novoselov, Nature
Photonics \textbf{6}, 749 (2012).
\bibitem{Koppens2011} F. H. L. Koppens, D. E. Chang, and F. J. Garca de
Abajo, Nano Letters \textbf{11}, 3370 (2011).
\bibitem{Tielrooij2015} K. J. Tielrooij, L. Orona, A. Ferrier, M. Badioli, G. Navickaite,
S. Coop, S. Nanot, B. Kalinic, T. Cesca, L. Gaudreau, Q. Ma, A. Centeno, A. Pesquera, A. Zurutuza,
H. de Riedmatten, P. Goldner, F. J. Garc\'{\i}a de Abajo,
P. Jarillo-Herrero, and F. H. L. Koppens, Nature Physics
\textbf{11}, 281 (2015).
\bibitem{Leggett1987} A. J. Leggett, S. Chakravarty, A. T. Dorsey, M. P. A.
Fisher, A. Garg, and W. Zwerger, Rev. Mod. Phys. \textbf{59},
1 (1987).
\bibitem{Orth2010} P. P. Orth, D. Roosen, W. Hofstetter, and K. Le Hur,
Phys. Rev. B \textbf{82}, 144423 (2010).
\bibitem{Winter2014} A. Winter and H. Rieger, Phys. Rev. B \textbf{90}, 224401 (2014).
\bibitem{Huttner1992} B. Huttner and S. M. Barnett, Phys. Rev. A \textbf{46}, 4306 (1992).
\bibitem{Drezet2017} A. Drezet, Phys. Rev. A \textbf{95}, 023831 (2017).
\bibitem{Philbin2010} T. G. Philbin, New Journal of Physics \textbf{12}, 123008 (2010).
\bibitem{Gruner1996} T. Gruner and D.-G. Welsch, Phys. Rev. A \textbf{53}, 1818
(1996).
\bibitem{Dung1998} H. T. Dung, L. Kn\"{o}ll, and D.-G. Welsch, Phys. Rev. A
\textbf{57}, 3931 (1998).
\bibitem{Edwards1975} S. F. Edwards and P. W. Anderson, Journal of Physics
F: Metal Physics \textbf{5}, 965 (1975).
\bibitem{Dzsotjan2010} D. Dzsotjan, A. S. S{\o}rensen, and M. Fleischhauer, Phys.
Rev. B \textbf{82}, 075427 (2010).
\bibitem{Kogut1979} J. B. Kogut, Rev. Mod. Phys. \textbf{51}, 659 (1979).
\bibitem{Ye1993} J. Ye, S. Sachdev, and N. Read, Phys. Rev. Lett. \textbf{70},
4011 (1993).
\bibitem{Kennett2001} M. P. Kennett, C. Chamon, and J. Ye, Phys. Rev. B \textbf{64},
224408 (2001).
\bibitem{Sachdev2011} S. Sachdev, “Quantum phase transitions,” (Cambridge University Press, 2011) Chap. 5.5.3, 2nd ed.
\bibitem{Aoki2014} H. Aoki, N. Tsuji, M. Eckstein, M. Kollar, T. Oka, and
P. Werner, Rev. Mod. Phys. \textbf{86}, 779 (2014).
\bibitem{sp} Supplemental Material.
\bibitem{Fischer1993} K. H. Fischer and J. A. Hertz, Spin Glasses, reprint
edition ed., Cambridge Studies in Magnetism (Book 1)
(Cambridge University Press, 1993).
\bibitem{Goto2008} H. Goto and K. Ichimura, Phys. Rev. A \textbf{77}, 053811
(2008).
\bibitem{Hanson2008} G. W. Hanson, J. Appl. Phys. \textbf{103}, 064302 (2008).
\bibitem{Tong2010} Q.-J. Tong, J.-H. An, H.-G. Luo, and C. H. Oh, Phys.
Rev. A \textbf{81}, 052330 (2010).
\bibitem{Yang2017} C.-J. Yang and J.-H. An, Phys. Rev. B \textbf{95}, 161408 (2017).
\bibitem{Thanopulos2017} I. Thanopulos, V. Yannopapas, and E. Paspalakis, Phys.
Rev. B \textbf{95}, 075412 (2017).
\bibitem{Gonzalez-Tudela2014} A. Gonz\'{a}lez-Tudela, P. A. Huidobro, L. Mart\'{\i}n-Moreno,
C. Tejedor, and F. J. Garc\'{\i}a-Vidal, Phys. Rev. B \textbf{89},
041402 (2014).
\bibitem{Gonzalez-Tudela2010} A. Gonz\'{a}lez-Tudela, F. J. Rodr\'{\i}guez, L. Quiroga, and
C. Tejedor, Phys. Rev. B \textbf{82}, 115334 (2010).
\bibitem{Dung2002} H. T. Dung, L. Kn\"{o}ll, and D.-G. Welsch, Phys. Rev. A
\textbf{66}, 063810 (2002).
\bibitem{Vega2017} I. de Vega and D. Alonso, Rev. Mod. Phys. \textbf{89}, 015001
(2017).
\bibitem{Stewart2008} J. T. Stewart, J. P. Gaebler, and D. S. Jin, Nature \textbf{454},
744 (2008).
\bibitem{Haussmann2009} R. Haussmann, M. Punk, and W. Zwerger, Phys. Rev.
A \textbf{80}, 063612 (2009).
\bibitem{Stratonovich1957} R. L. Stratonovich, Soviet Physics Doklady \textbf{2}, 416 (1957).
\bibitem{Hubbard1959} J. Hubbard, Phys. Rev. Lett. \textbf{3}, 77 (1959).
\end{thebibliography}
\end{document}